\documentclass{jfm}
\usepackage{amsmath,amsfonts,amssymb,upmath,graphicx}
\newcommand{\ud}{\mathrm{d}}
\newcommand{\ue}{\mathrm{e}}
\newcommand{\ui}{\mathrm{i}}
\title{Does multifractal theory of turbulence have logarithms in the scaling relations?}
\author[U. Frisch, M. Martins Afonso, A. Mazzino and V. Yakhot]{U.\ns F\ls R\ls I\ls S\ls C\ls H$^1$,\ns M.\ns M\ls A\ls R\ls T\ls I\ls N\ls S\ns A\ls F\ls O\ls N\ls S\ls O$^2$,\newline
A.\ns M\ls A\ls Z\ls Z\ls I\ls N\ls O$^2$ \and V.\ns Y\ls A\ls K\ls H\ls O\ls T$^3$}
\affiliation{$^1$CNRS,~Lab.~Cassiop\'ee,~Observatoire~de~la~C\^ote~d'Azur,~B.P.~4229,~06304~Nice~Cedex~4,~France
\\[\affilskip]
$^2$INFM-Dipartimento~di~Fisica,~Universit\`a~di~Genova
and~Istituto~Nazionale~di~Fisica~Nucleare,~Sezione~di~Genova,
Via~Dodecaneso~33,~16146~Genova,~Italy
\\[\affilskip]
$^3$Department of Aerospace and Mechanical Engineering, Boston University, 02215 Boston, USA}
\date{\today}

\usepackage{natbib}
\allowdisplaybreaks[1]

\begin{document}

\maketitle

\begin{abstract}

 The multifractal theory of turbulence uses a saddle-point evaluation in
 determining the power-law behaviour of structure functions. Without suitable
 precautions, this could lead to the presence of logarithmic corrections,
 thereby violating known exact relations such as the four-fifths law. Using
 the theory of large deviations applied to the random multiplicative model of
 turbulence and calculating subdominant terms, we explain here why such
 corrections cannot be present.

\end{abstract}

\section{Introduction}

 In fully developed turbulence there is now fairly good evidence for
 anomalous scaling, that is scaling exponents which cannot be predicted
 by dimensional analysis. Some of this evidence is reviewed in
 \cite{F95}. This reference also contains a detailed presentation of the
 multifractal formalism in the formulation of \cite{PF85} (henceforth PF; see
 also \cite{BPPV84}).
 In this formalism, anomalous scaling for structure functions (moments
 of velocity increments) is connected by a Legendre transformation
 to the distribution of singularities
 of the velocity field. An earlier and alternative formalism for
 anomalous scaling was introduced by the Russian School of
 Kolmogorov \cite[][]{O62,K62,Y66}. In its simplest version it
 uses a random multiplicative model for calculating the statistical
 fluctuations of the energy dissipation on various scales; the
 fractal properties of this model were discovered by
 \cite{M74}. The bridging of the two formalisms is discussed
 in \cite{F95} in the light of the theory
 of large deviations for the sums of independent identically distributed
 random variables, discovered in the thirties by \cite{C38}.

 We recall that in its original formulation PF gives an integral
 representation of the structure functions which are then evaluated by the
 method of steepest descent through a saddle point. Specifically,
 the structure function of order $p$ for a separation $\ell$ is given by
 \begin{equation} \label{dmuh}
  \frac{S_p(\ell)}{v_0^p}\sim\int\!\ud\mu(h)\,\left(\frac{\ell}{\ell_0}\right)^{ph+3-D(h)}\;.
 \end{equation}
 We have here used the notation of \cite{F95}: $v_0$ is the r.m.s.\ velocity
 fluctuation, $\ell_0$ is the integral scale, $D(h)$ is the fractal dimension
 associated to singularities of scaling exponent $h$ and $\ud\mu(h)$ gives the
 weight of the different exponents For a given $p$, let us assume that
 the exponent
 $ph+3-D(h)$ has a minimum $\zeta_p$, as a function of $h$ and that it behaves
 quadratically near this minimum. Standard application of Laplace's method
 of steepest descent \cite[see, e.g.,][]{BO78} shows then that, at small
 separations, $S_p(\ell)$ varies as $(\ell/\ell_0)^{\zeta_p}$
 but \emph{with a logarithmic prefactor} $[\ln(\ell/\ell_0)]^{-1/2}$ stemming
 from the Gaussian integration near the minimum. For $p=3$ this logarithmic
 prefactor is clearly inconsistent with the
 four-fifths law of \cite{K41}, one of the very few exact results in
 high-Reynolds number turbulence, which tells us that the third-order
 (longitudinal) structure function is given by
 $-(4/5)\varepsilon\ell$, where $\varepsilon$ is the mean energy dissipation 
 per unit mass. 

 In \cite{F95} this difficulty was handled by writing
 \begin{equation} \label{asymptstrcorrect}
  \lim_{\ell\to0}\frac{\ln S_p(\ell)}{\ln\ell}=\zeta_p\;.
  \end{equation}
 Indeed, by taking the logarithm of the structure function we change the
 \emph{multiplicative} logarithmic correction into an \emph{additive} log-log
 correction which, after division by $\ln\ell$, becomes \emph{subdominant} as
 $\ell\to0$. But if we do not take the logarithm of the structure function, is
 there a logarithmic correction in the leading term whose presence, for $p=3$,
 would invalidate the standard multifractal formalism? 

 It is also well known that such logarithmic corrections are absent in the
 random multiplicative model, a matter we shall come back to. Actually, as we
 shall see, the random multiplicative model gives the key allowing us to
 understand why logarithmic corrections are unlikely and are definitely ruled 
 out in the third-order structure function.

 In \S~\ref{sec:mmm} we recall some basic facts about the random
 multiplicative model and the way it can be reformulated in terms of
 multifractal singularities using large-deviations theory. This is the theory
 introduced by \cite{C38} which allows the estimation of the (very small)
 probability that the sum of a large number of random variables deviates
 strongly from the law of large numbers.\footnote{The reader is only assumed
 to be familiar with a very elementary version of large-deviations theory, as
 explained e.g. in Sec.~8.6.4. of \cite{F95}; a more detailed exposition for
 physicists interested e.g. in the foundations of thermodynamics can be found
 in \cite{L73}; \cite{V84} and \cite{DZ98} are written for the more
 mathematically minded reader.} In this way the random multiplicative model
 can be reformulated in standard multifractal language; naive application of
 PF would then suggest the presence of logarithmic corrections. In
 \S~\ref{sec:rld} we show how a refined version of large-deviations theory,
 which goes beyond leading order, removes the logarithmic corrections, which
 are cancelled by other logarithmic corrections in the probability densities.
 In \S~\ref{sec:end} we summarise our findings, return to the general
 multifractal formalism beyond the specific random multiplicative model and
 show that the four-fifths law allows us to obtain the first subleading
 correction to the usual multifractal probability.

\section{The random multiplicative model} 
\label{sec:mmm}

 In the random multiplicative model \cite[see, e.g.,][for details]{F95} one
 assumes that an integral-scale size cube with side $\ell_0$ is subdivided
 into 8 first-level cubes of half the side, which in turn are divided into
 $8^2$ second-level cubes of side $\ell_02^{-2}$, and so on. The ``local''
 dissipation at the $n$th level with scale $\ell=\ell_0 2^{-n}$ is defined as
 \begin{equation} \label{multw}
  \varepsilon_\ell=\varepsilon W_1W_2\ldots W_n,
 \end{equation}
 where $\varepsilon$ is a non-random mean dissipation per unit mass and the
 $W_i$'s are positive, independently and identically distributed random
 variables of unit mean value.\\ The ensemble average
 $\langle\varepsilon_\ell\rangle$ is thus still equal to $\varepsilon$, so
 that the cascade is conservative only in the mean. Since we are interested
 in describing the inertial-range scaling properties ($\ell\ll \ell_0$), we
 shall mainly focus our attention on high-order generations, i.e.\ on large
 values of $n$, where large fluctuations of $\varepsilon_\ell$ are present. As
 a consequence, the formalism of multiplicative variables leads to the
 presence of very large fluctuations. The correspondence between
 multifractality and the probabilistic theory is expressed by the relationship
 \begin{equation} \label{rp}
  n=-\log_2\frac{\ell}{\ell_0}=-\frac{1}{\ln2}\ln\frac{\ell}{\ell_0}\;.
 \end{equation}
 Thus $n$ can be viewed either as the number of $W$ factors determining
 the local dissipation or as the number of cascade steps leading from
 the ``injection'' length scale $\ell_0$ to the current scale $\ell$.

 It is elementary and standard to show that the moments of the
 local dissipation are given by
 \begin{equation} \label{tp}
  T_p(\ell)\equiv\langle\varepsilon_l^p\rangle=\langle(\varepsilon W_1\cdots W_n)^p\rangle=\varepsilon^p\langle W^p\rangle^n=\varepsilon^p\left(\frac{\ell}{\ell_0}\right)^{-\log_2\langle W^p\rangle}\;.
 \end{equation}

 Then, following the suggestion originally made
 by \cite{O62}, one calculates the structure functions at
 separation $\ell$ by the \cite{K41} expression in which one
 replaces the mean dissipation by its local random value
 $\varepsilon_\ell$, to obtain
 \begin{equation} \label{fs}
  S_p(\ell)\approx\langle(\ell\varepsilon_\ell)^{p/3}\rangle=l^{p/3}T_{p/3}(\ell)\propto\ell^{\zeta_p}\;,
 \end{equation}
 where $\zeta_p=p/3-\log_2\langle W^{p/3}\rangle$. Obviously the
 third-order
 structure function has exponent unity, as required by the four-fifths
 law and none of the structure functions has any multiplicative
 logarithmic factor. 

 There is however an alternative and somewhat roundabout way of
 evaluating the structure functions for the random multiplicative
 model. First we transform the product of positive random variables
 into a sum by setting $W_i = 2^{-m_i}$, to obtain
 \begin{eqnarray}
  T_p(\ell)&=&\langle(\varepsilon2^{-m_1}\cdots2^{-m_n})^p\rangle=\varepsilon^p\langle2^{-m_1p}\cdots2^{-m_np}\rangle \nonumber \\
  &=&\varepsilon^p\langle2^{-nxp}\rangle=\varepsilon^p\int\!\ud x\,\ue^{-nxp\ln2}P_n(x)\;, \label{TP}
 \end{eqnarray}
 where
 \begin{equation} \label{va}
  x\equiv\frac{m_1+\ldots+m_n}{n}
 \end{equation}
 is the sample mean of $n$ independent and identically
 distributed random variables and $P_n(x)$ its probability density
 function (PDF). When the PDF $P(m)$ 
 of the individual variables $m_i$ falls off
 very quickly at large arguments, as is usually assumed in the
 random multiplicative model,\footnote{We shall here assume that $P(m)$
 falls off faster than exponentially at large $|m|$.} all the moments of the $m_i$s are finite 
 and the law of large numbers \cite[][]{F68} implies that $P_n(x)$ is,
 for large $n$, increasingly concentrated near the mean $\langle m\rangle$.
 The theory of large deviations \cite[][]{C38} tells us
 roughly that when $x\neq\langle m\rangle$ its probability falls
 off exponentially with $n$ as $\ue^{ns(x)}$, where the Cram\'er function
 $s(x)$ is non-positive up-convex and vanishes at $x=\langle m\rangle$. The
 Cram\'er function can be expressed as the Legendre transform 
 \begin{equation} \label{legendre}
  s(x)=\inf_{\alpha}\,[\alpha x+\ln Z(\alpha)]
 \end{equation}
 of the characteristic function
 \begin{equation} \label{defchar}
  Z(\alpha)\equiv\int\!\ud y\,\ue^{-\alpha y}P(y)=\langle\ue^{-\alpha m}\rangle\;.
 \end{equation}
 The correct statement of large deviations is that $\ln P_n(x)/n$ tends
 to $s(x)$ as $n\to\infty$. Suppose however we somewhat sloppily write the
 large-deviations result as 
 \begin{equation} \label{incorrect}
  P_n(x)\sim\ue^{ns(x)}\; \quad \hbox{INCORRECT}
 \end{equation}
 and use this in (\ref{TP}). We then obtain an integral representation for
 $T_p(\ell)$ and thus for $S_p(\ell)$ which when evaluated by steepest
 descent for large $n$ will give not just a power law in $\ell$ but
 also a multiplicative correction proportional to
 $1/\sqrt{-\ln(\ell/\ell_0)}$,
 thus contradicting
 (\ref{fs}). 

 To resolve the paradox we need to extend the large-deviations result
 beyond the leading order. This is called the theory of refined large
 deviations, first developed by \cite{BRR60} and which is reviewed in 
 \cite{DZ98}. In the next section we shall show how this can be
 done by rather elementary application of steepest descent.

\section{Refined large-deviations theory and the disappearance of logs} \label{sec:rld}

\begin{figure} \label{f:Candsaddle}
  \includegraphics[width=10cm]{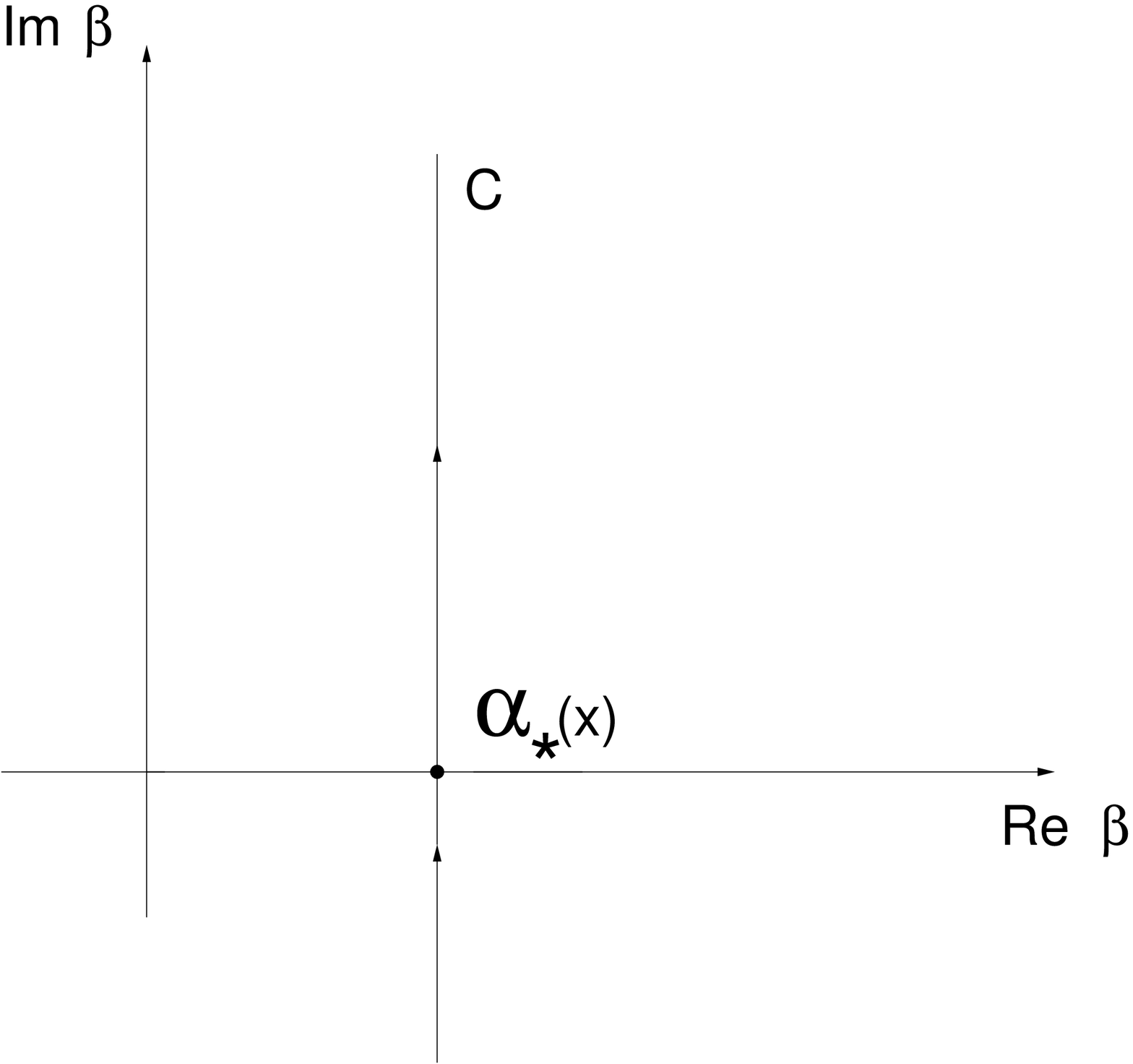}
  \caption{The integration contour $C$ through the saddle $\alpha_\star$ in the complex $\beta$-plane.}
 \end{figure}

 We now derive the asymptotic expansion for large $n$ of the PDF
 $P_n(x)$ of the sample mean (\ref{va}). Consider the characteristic
 function of the sample mean
 \begin{eqnarray}
  Z_n(\alpha)&\equiv&\int\!\ud x\,\ue^{-\alpha x}P_n(x)=\langle\ue^{-\alpha(m_1+\ldots+m_n)/n}\rangle \label{laplace} \\
  &=&\langle\ue^{-\alpha m_1/n}\cdots\ue^{-\alpha m_n/n}\rangle=\langle\ue^{-\alpha m/n}\rangle^n=Z^n\left(\frac{\alpha}{n}\right)\;, \label{zntoz}
 \end{eqnarray}
 where $Z(\alpha)$ is the characteristic function for a single
 variable $m$, defined in (\ref{defchar}). Since we assumed that $P(m)$
 falls
 off faster than exponentially, $Z(\alpha)$ can be defined for any
 complex $\alpha$. Hence we can invert the Laplace transform appearing
 in (\ref{laplace}) by a Fourier integral along a contour $C$ running
 from $-\ui\infty$ to $+\ui\infty$ (see Fig.~\ref{f:Candsaddle})
 \begin{equation} \label{tl}
  P_n(x)=\frac{1}{2\ui\upi}\int_C\!\ud\beta\,\ue^{\beta x}Z^n\left(\frac{\beta}{n}\right)\;.
 \end{equation}
 We recast (\ref{tl}) in exponential form
 \begin{equation} \label{cv}
  P_n(x)=\frac{1}{2\ui\upi}\int_C\!\ud\beta\,\ue^{\beta x+n\ln Z(\beta/n)}=\frac{n}{2\ui\upi}\int_C\!\ud\gamma\,\ue^{n[\gamma x+\ln Z(\gamma)]}\;,
 \end{equation}
 with the substitution $\gamma=\beta/n$. By (\ref{legendre}) the argument
 of the exponential has a minimum $s(x)$ along the real $\gamma$-line at a
 point $\alpha_\star(x)$. Taking the contour $C$ through $\alpha_\star(x)$,
 the argument of the exponential will now have a maximum at this point;
 thus (\ref{cv}) can be evaluated by steepest descent 
 \cite[see, e.g.,][]{BO78}.

 We recall that for an integral of the form 
 \begin{equation} \label{in}
  I(n)=\int_C\!\ud y\,f(y)\ue^{n\phi(y)}\;,
 \end{equation}
 with a saddle point $y_{\star}$ where $\phi'$ vanishes and
 where neither $f$ nor $\phi''$ vanish,
 the large-$n$ behaviour is given by 
 \begin{equation} \label{ps}
  I(n)=\sqrt{\frac{2\upi}{-n\phi''(y_{\star})}}\,\ue^{n\phi(y_{\star})}f(y_{\star})\left[1+O\left(\frac{1}{n}\right)\right]\;.
 \end{equation}
 
 The saddle point formula (\ref{ps}), applied to (\ref{cv}) gives, after
 taking a logarithm
 \begin{equation} \label{gd}
  \frac{\ln P_n(x)}{n}=s(x)+\frac{\ln n}{2n}-\frac{\ln\left(2\upi\,Q\right)}{2n}+O\left(\frac{1}{n^2}\right)\;,
 \end{equation}
 where $Q>0$ is the second derivative of $\alpha x+\ln Z(\alpha)$,
 evaluated at the saddle point $\alpha_{\star}(x)$. 
 As $n$ does not 
 appear in $Q$, which is solely a function of $x$,
 the right-hand side of (\ref{gd}) is thus structured as an inverse
 power series in $n$, except for the first subleading term which
 contains a logarithm.

 Note that expressions such as (\ref{gd}) are very common in thermodynamic
 applications of large deviations when dealing with the logarithm of the (very
 large) number of states \cite[see, e.g.,][]{L73}.

 We can of course rewrite (\ref{gd}) in exponentiated form as
 \begin{equation} \label{gdexp}
  P_n(x)=\sqrt{n/(2\upi\,Q)}\,\ue^{ns(x)}\left[1+O\left(\frac{1}{n}\right)\right]\;.
 \end{equation}
 In this way we see that $P_n(x)$ has a multiplicative $\sqrt{n}$
 correction. Recalling that in the random multiplicative model
 $n=-\log_2(\ell/\ell_0)$, this correction is just what we need to
 cancel the unwanted logarithms in the structure functions obtained
 when the incorrect form (\ref{incorrect}) is used. 

 It is important to note that the quantity which goes to a finite
 limit for large $n$ is $\ln P_n(x)/n$ and that for this quantity
 the correction we have determined is an \emph{additive subleading
 term}. This is why such terms should be regarded as subleading.
 Of course the cancellation of logarithms for the random multiplicative
 model cannot take place just at the first subleading order, since 
 (\ref{fs}) is an exact expression and has no logarithms.
 To evaluate refined large deviations to all orders for a general
 random multiplicative model is quite cumbersome and we shall not
 attempt it here because it would not shed additional light on the issue
 discussed. It can however be done quite easily for simple random
 multiplicative models such as the black-and-white model of \cite{NS64}.

\section{Back to multifractal turbulence}
\label{sec:end}

 In multifractal language, the result obtained within the framework of
 the random multiplicative model is that the probability $P(\ell,h)$
 to be within
 a distance $\ell$ of the set carrying singularities of scaling
 exponent between $h$ and $h+\ud h$ is not
 $(\ell/\ell_0)^{3-D(h)}\ud\mu(h)$ but is
 actually given, for small $\ell$, by
 \begin{equation} \label{better}
  P(\ell,h)\propto\left(\frac{\ell}{\ell_0}\right)^{3-D(h)}\left[-\ln\frac{\ell}{\ell_0}\right]^{1/2}\ud\mu(h)\;,
 \end{equation}
 which has a subleading logarithmic correction.
 We recall that it must be
 qualified ``subleading'' because the correct statement of the large-deviations
 leading-order result involves the logarithm of the probability divided
 by the logarithm of the scale. The correction is then a subleading additive 
 term.

 It is important to mention that the presence of a square root of a logarithm
 correction in the multifractal probability density has already been
 proposed by \cite{MS89} on the basis of a normalization requirement;
 they observed that without such a correction the singularity spectrum
 $f(\alpha)$ comes out wrong; they also pointed out that a similar correction
 has been proposed by \cite{vdWS88} in connection with the measurement of
 generalized Renyi dimensions. 

 Returning to the multifractal formalism of turbulence, beyond the
 random multiplicative model, we observe that the usual multifractal
 ansatz as made in \cite{PF85} is only about the leading term of the 
 probability, which is easily reinterpreted in geometrical language.
 Hence, it does not allow us to determine logarithmic corrections
 in structure functions. However, if we use Kolmogorov's four-fifths
 law, we have an additional piece of information which implies
 that the multifractal probability should have a subleading
 logarithmic correction with precisely the form it has in
 (\ref{better}). This improved form then rules out subleading logarithmic 
 corrections in any of the structure functions.

\begin{acknowledgements}

 We wish to thank J.~Bec, M.~Blank, G.~Boffetta, A.~Celani, A.~Dembo,
 G.~Eyink, C.~Meneveau, K.R.~Sreenivasan and A.~Vulpiani for illuminating discussions and
 suggestions. This research has been partially supported by the MIUR under
 contract Cofin.~2003 (prot. 020302002038).

\end{acknowledgements}

\end{document}